\documentclass[A4paper,unsortedaddress,superscriptaddress,prl,twocolumn]{revtex4-1}
\usepackage{hyperref}
\usepackage{epsfig}
\usepackage{graphicx}
\usepackage{subfigure}
\usepackage{latexsym}
\usepackage{color}
\usepackage{fullpage}
\usepackage{dcolumn}
\usepackage{bm}
\usepackage{ulem}
\usepackage{units}

\begin{document}

\title{Cobalt intercalation at the graphene/iridium(111) interface: influence of rotational domains, wrinkles and atomic steps}

\author{S. Vlaic$^{1,2}$, A. Kimouche$^{1,2}$, J. Coraux$^{1,2}$, B. Santos$^3$, A. Locatelli$^3$, N. Rougemaille$^{1,2}$}

\address{$^1$ CNRS, Inst NEEL, F-38042 Grenoble, France\\$^2$ Univ. Grenoble Alpes, Inst NEEL, F-38042 Grenoble, France\\$^3$ Elettra - Sincrotrone Trieste S.C.p.A., S.S: 14 km 163.5 in AREA Science Park, I-34149 Basovizza, Trieste, Italy}

\begin{abstract}
Using low-energy electron microscopy, we study Co intercalation under graphene grown on Ir(111). Depending on the rotational domain of graphene on which it is deposited, Co is found intercalated at different locations. While intercalated Co is observed preferentially at the substrate step edges below certain rotational domains, it is mostly found close to wrinkles below other domains. These results indicate that curved regions (near substrate atomic steps and wrinkles) of the graphene sheet facilitate Co intercalation and suggest that the strength of the graphene/Ir interaction determines which pathway is energetically more favorable.
\end{abstract}


\maketitle

In view of potential technological appications, the ability to modify and control the properties of a graphene layer has been a central issue since its discovery\cite{Novoselov2012}. The intercalation of foreign atoms or molecules between a graphene sheet and its substrate often affects the electronic and magnetic properties of the considered interface. For example, the intercalation of noble metals \cite{Shikin2000,Dedkov2001} or hydrogen atoms \cite{Riedl2009} can be used to reduce the interaction between graphene and its substrate, and even restore the electronic properties of free-standing graphene, while the intercalation of alkali metals is an efficient mean to control the doping level of graphene \cite{Nagashima1994}. Intercalation of a ferromagnetic transition metal can also enhance the net magnetic moment induced in carbon atoms when graphene is in contact with a magnetic surface \cite{Weser2011}, and is a promising route to fabricate graphene/ferromagnetic metal hybrid structures with perpendicular magnetic anisotropy \cite{Rougemaille2012, Decker2013}.

\begin{figure}
\begin{center}
\includegraphics[width=8cm]{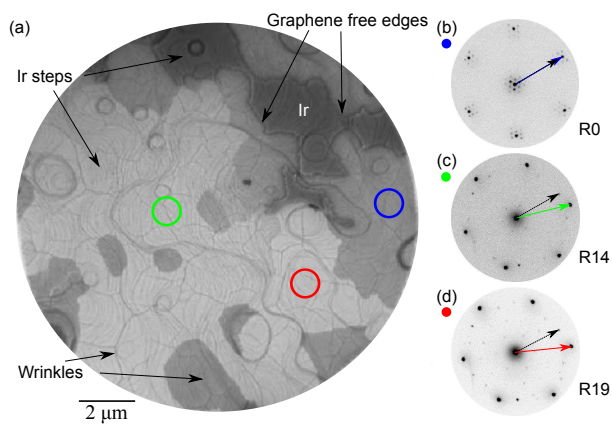}
\caption{(a) 15 $\mu$m field of view LEEM image ($V_{start}=15$ V) of the graphene/Ir(111) surface, showing bare Ir (darkest regions) and three graphene rotational domains. Thin lines all over the surface are Ir atomic steps, while thick lines on the graphene-covered surface are wrinkles. (b)-(d) 2 $\mu$m-diameter selected area of the LEEM image where low-energy electron diffraction has been performed to identify the three graphene rotational domains. In these patterns ($V_{start}=40$ V), the black dash arrows indicate the Ir diffraction peaks, while the blue, green and red arrows indicate the carbon peaks for R0, R14 and R19 domains, respectively.}
\end{center}
\end{figure}

Understanding where and how a foreign species intercalates below graphene is a challenging task, and different scenarios have been proposed. While oxygen intercalates at the free edges of graphene grown on Ru(0001) \cite{Sutter2010, Starodub2010} and on Ir(111) \cite{Granas2012}, alkali metals instead may intercalate at the substrate step edges or at boundaries between different rotational domains in graphene/Ni(111) \cite{Nagashima1994} and in graphite \cite{Wu1983}. Regarding transition metals, the intercalation mechanism remains elusive. While it has been demonstrated that pre-existing defects in graphene, such as vacancies or pentagon-heptagon pairs, reduce the required energy to trigger intercalation \cite{Coraux2012, Sicot2012}, several recent experimental works have shown that other mechanisms could be at work. In particular, the formation of atomic defects, not pre-existing in the graphene layer but induced by the contact with a transition metal cluster, with subsequent restoring of the carbon-carbon bonds, has been suggested as a possible way for metal intercalation \cite{Sicot2012, Huang2011,Jin2013}.

In this work, low-energy electron microscopy\cite{Bauer1994, Altman2010} (LEEM) is used to study cobalt intercalation at moderate annealing temperature (about 125$^\circ$C) underneath graphene grown on an iridium (111) surface. Depending on the rotational orientation of the graphene domain, we find Co intercalated at different locations. Below R0 domains, where the carbon zigzag rows are aligned with the Ir dense-packed atomic rows, intercalated Co is observed at the Ir atomic steps. Below other rotational domains, intercalated Co is mostly found close to graphene wrinkles, which correspond to local, linear delaminated regions of the graphene sheet induced by strain relief upon cooling the sample after high temperature growth \cite{Alpha2009,Hattab2012}. 

\begin{figure}
\begin{center}
\includegraphics[width=6.5cm]{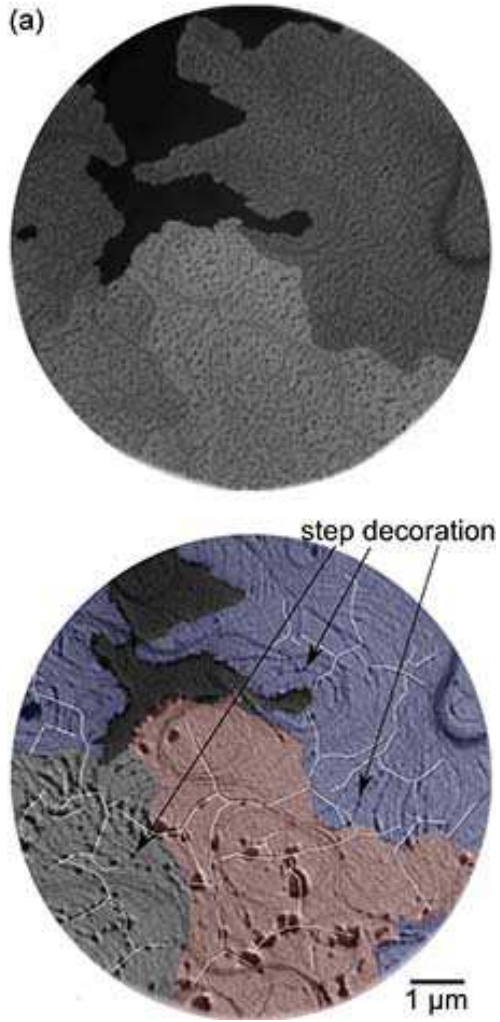}
\caption{(a) 6 $\mu$m field of view LEEM image ($V_{start}=10$ V) of the graphene/Ir(111) surface after deposition at room temperature of a 1 ML thick Co layer, which results in randomly distributed clusters (dark dots) atop the graphene sheet. (b) LEEM image ($V_{start}=4$ V) of the same region of the surface after annealing at 125$^\circ$C. The different rotational domains have been colored for clarity. R0, R14 and R19 appear blue, grey and red, respectively. Wrinkles are highlighted by white lines.}
\end{center}
\end{figure}

The LEEM measurements were performed at the Nanospectroscopy beamline of the Elettra synchrotron radiation facility \cite{Locatelli2006}. Samples were prepared in situ in ultra-high vacuum (UHV) conditions (base pressure 5$\times$10$^{-11}$ mbar). An Ir(111) single crystal was used as a substrate and cleaned with repeated cycles of Ar ion sputtering and high temperature (1200$^\circ$C) flashes under oxygen (10$^{-8}$ mbar). A last temperature flash (1200$^\circ$C) is finally done under UHV conditions to remove the oxide layer. Graphene was obtained by chemical vapor deposition, by exposing the Ir(111) surface to 5$\times$10$^{-8}$ mbar of C$_2$H$_4$ at 1000$^\circ$C. In all experiments presented here, graphene covers about 80\% of the Ir surface. Co was subsequently deposited at room temperature from an electron-beam evaporation source at a rate of about 0.3 monolayer (ML) per minute \cite{mono}. The surface is then imaged while annealing.

The LEEM image in Fig. 1(a) illustrates the typical aspect of the graphene/Ir(111) surface, prior to Co deposition. In this image, bare Ir appears as the darkest regions. Consistent with previous literature \cite{Loginova2009, Meng2012}, we observe several rotational domains across the graphene layer with different LEEM contrasts [see different shades of grey in Fig. 1(a)]. Microdiffraction (LEED) measurements are used to identify the local crystallographic orientation of these rotational domains, as shown in Figs. 1(b)-(d). Besides R0, these orientations correspond to rotation of 14$^\circ$ and 19$^\circ$ of the carbon zigzag rows with respect to the Ir dense-packed atomic rows. In the following, we refer to R14 and R19 when describing these rotational domains. In other places of the surface, we also observe R30 domains (not shown here). Thin lines on the surface are the substrate atomic steps, while thick lines are wrinkles [see Fig. 1(a)], as previously observed\cite{Alpha2009}.

In agreement with other works \cite{Vovan2011}, deposition of 1 ML of Co at room temperature leads to the formation of randomly distributed clusters all over the graphene surface [see black dots in Fig. 2(a)]. Annealing of the Co ML at about 125$^\circ$C strongly modifies the surface morphology, as shown in Fig. 2(b). We then observe extended dark areas below R14 and R19 where Co accumulates, while, surprisingly, the surface morphology below R0 remains essentially unchanged, although Co starts decorating the Ir atomic steps [the LEEM contrast becomes darker along the substrate step edges below R0 domains, see Fig. 2(b)]. In the following, we show that these dark regions correspond to intercalated Co.

To do so, we use low-energy electron reflectivity\cite{Bauer1994} to measure the relative changes of the surface work function (WF) at different steps of the sample preparation. Here, we define the relative work function as the energy for which electron reflectivity drops by 10\% from total reflectance \cite{Starodub2011}. Following previous works \cite{Rougemaille2012, Coraux2012} to track Co intercalation, we take advantage of the fact that the WF of graphene-terminated metal surfaces is often lower than the WF of the corresponding clean metal surfaces \cite{Giovannetti2008}. This is particularly true for Co and Ir: $\delta$WF$_{Co}$=WF$_{Co}$-WF$_{Gr/Co}\sim$1.7 eV\cite{Giovannetti2008} and  $\delta$WF$_{Ir}$=WF$_{Ir}$-WF$_{Gr/Ir}\sim$1 eV \cite{Loginova2009,Starodub2011}.

\begin{figure}
\begin{center}
\includegraphics[width=8cm]{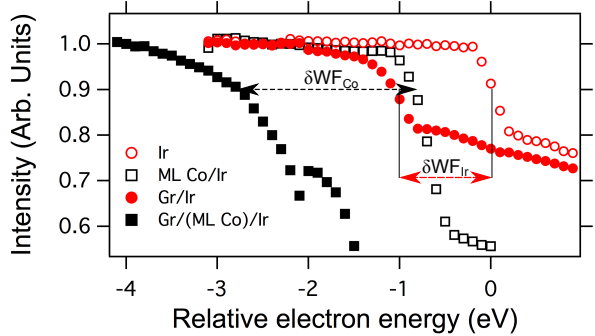}
\caption{Work function measurement of clean Ir (taken as a reference) and graphene/Ir surfaces (empty and full red circles, respectively). The same comparison is made between 1 ML of Co on Ir (empty squares) and the dark regions observed on the R14 and R19 domains (filled squares).}
\end{center}
\end{figure}

We thus compare the surface WF of clean Ir with the one of graphene/Ir and the WF of 1 ML of Co on Ir with the one measured on the dark regions in Fig. 2(b). The results, reported in Fig. 3, show that $\delta$WF$_{Ir}\sim$1 eV, in agreement with previous experimental works \cite{Loginova2009, Starodub2011}, and that $\delta$WF$_{Co}\sim$1.8 eV, demonstrating that the dark regions in Fig. 2(b) are indeed intercalated Co. Note that, within our 100 meV energy resolution, the surface WF of the graphene covered regions does not change upon Co deposition nor after 125$^\circ$C annealing, except for the dark regions in Fig. 2(b)\cite{supp}.
Interestingly, the amount of intercalated Co seems significantly different for the three rotational domains. For example, while intercalated Co makes relatively large clusters below the R14 and R19 domains, it is hardly detectable below R0 [see Fig. 2(b)]. This result suggests that the graphene rotational domains play a role in the intercalation process: stronger graphene/Ir interaction (R0 domains) impede Co intercalation at 125$^\circ$C, while it is already activated at domains (R14 and R19) where this interaction is weaker\cite{Loginova2009}.

The LEEM image shown in Fig. 2(b) reveals another intriguing aspect: while intercalated Co is preferentially observed at Ir step edges below R0 domains, the intercalated regions below R14 and R19 often appear close to graphene wrinkles. This is striking in Figure 2(b): regions where Co is intercalated are strongly correlated to the positions of graphene wrinkles below R14 and R19 domains, contrary to what is observed below R0 domains. Part of these results are consistent with recent STM measurements in which Co\cite{Decker2013} and Ni\cite{Pacile2013} islands intercalated at the graphene/Ir(111) interface have been found mostly at the Ir step edges under the R0 rotational domain. However, the role of graphene wrinkles and rotational domains in the intercalation process of a transition metal has not been anticipated in these earlier studies. 

One may think that pre-existing defects in the graphene layer, located at the substrate step edges and wrinkles, facilitate Co intercalation. However, this is unlikely as there is no reason to have pre-existing defects only close to atomic steps at R0 domains, and only close to wrinkles at the other rotational domains. In other words, if pre-existing defects were the key ingredient to explain our findings, we would expect Co intercalation both at substrate step edges and wrinkles, independently of the rotational domains. This is not what we find. Besides, intercalation has not been observed at domain boundaries, i.e. between two different rotational domains, where pentagon-heptagon defects are located \cite{Coraux2008}. This suggests, in the temperature window we probe in this work, that curved regions (substrate atomic steps and wrinkles) of the graphene sheet are preferential sites for Co intercalation compared to flatter regions, such as domain boundaries\cite{comment}. Finally, we find only negligible amount of intercalated Co at the free edges of the graphene layer. This suggests that the energy barriers for intercalation are lower when Co is on top of graphene than on the Ir surface, where graphene-Ir binding is strong\cite{Loginova2009b, Locatelli2013}. 

For flat and free standing graphene, the formation of a vacancy requires high energy, about 7 eV \cite{Thrower1978}, but this value drops to 1.5 eV in the presence of Co clusters \cite{Boukhvalov2009}. In addition, previous works on carbon nanotubes showed that the vacancy formation energy decreases with increasing curvature \cite{Krasheninnikov2006}. In particular, for a nanotube diameter of 5 \AA, corresponding approximately to the graphene curvature at the Ir step edges \cite{Coraux2008}, the vacancy formation energy is reduced by 2 eV with respect to free standing graphene \cite{Krasheninnikov2006}. The di-vacancy formation energy is even smaller both in presence of Co clusters and for curved graphene sheets. It is then reasonable to believe that defects form in the graphene sheet when a Co cluster has nucleated at a graphene wrinkle or at an Ir atomic step, without the need of pre-existing defects.

In summary, Co is found intercalated at regions where graphene has a strong curvature, across the substrate step edges (below R0 domains) and where it is wrinkled (below the other domains). These results suggest that the strength of the graphene/Ir interaction determines which pathway is energetically more favorable. Our work opens opportunities to prepare systems where the intercalation mechanism could be controlled at the nanoscale. Moreover, the pathways we unveiled are already operational at moderate annealing temperature, at which intermixing between the intercalated metal and the substrate is negligible.

The authors thank X. Blase, L. Magaud, C. Chapelier and C. Tonnoir for fruitful discussions and T.O. Mente\c{s} for careful reading of the manuscript. S.V. acknowledges support by the Swiss National Science Foundation through project PBELP2-146587 and J.C. acknowledges support from the EU GRENADA and the ANR NANOCELLS projects.


\begin{thebibliography}{35}%
\makeatletter
\providecommand \@ifxundefined [1]{%
 \@ifx{#1\undefined}
}%
\providecommand \@ifnum [1]{%
 \ifnum #1\expandafter \@firstoftwo
 \else \expandafter \@secondoftwo
 \fi
}%
\providecommand \@ifx [1]{%
 \ifx #1\expandafter \@firstoftwo
 \else \expandafter \@secondoftwo
 \fi
}%
\providecommand \natexlab [1]{#1}%
\providecommand \enquote  [1]{``#1''}%
\providecommand \bibnamefont  [1]{#1}%
\providecommand \bibfnamefont [1]{#1}%
\providecommand \citenamefont [1]{#1}%
\providecommand \href@noop [0]{\@secondoftwo}%
\providecommand \href [0]{\begingroup \@sanitize@url \@href}%
\providecommand \@href[1]{\@@startlink{#1}\@@href}%
\providecommand \@@href[1]{\endgroup#1\@@endlink}%
\providecommand \@sanitize@url [0]{\catcode `\\12\catcode `\$12\catcode
  `\&12\catcode `\#12\catcode `\^12\catcode `\_12\catcode `\%12\relax}%
\providecommand \@@startlink[1]{}%
\providecommand \@@endlink[0]{}%
\providecommand \url  [0]{\begingroup\@sanitize@url \@url }%
\providecommand \@url [1]{\endgroup\@href {#1}{\urlprefix }}%
\providecommand \urlprefix  [0]{URL }%
\providecommand \Eprint [0]{\href }%
\providecommand \doibase [0]{http://dx.doi.org/}%
\providecommand \selectlanguage [0]{\@gobble}%
\providecommand \bibinfo  [0]{\@secondoftwo}%
\providecommand \bibfield  [0]{\@secondoftwo}%
\providecommand \translation [1]{[#1]}%
\providecommand \BibitemOpen [0]{}%
\providecommand \bibitemStop [0]{}%
\providecommand \bibitemNoStop [0]{.\EOS\space}%
\providecommand \EOS [0]{\spacefactor3000\relax}%
\providecommand \BibitemShut  [1]{\csname bibitem#1\endcsname}%
\let\auto@bib@innerbib\@empty
\bibitem [{\citenamefont {Novoselov}\ \emph {et~al.}(2012)\citenamefont
  {Novoselov}, \citenamefont {Fal'ko}, \citenamefont {Colombo}, \citenamefont
  {FGellert}, \citenamefont {Schwab},\ and\ \citenamefont
  {Kim}}]{Novoselov2012}%
  \BibitemOpen
  \bibfield  {author} {\bibinfo {author} {\bibfnamefont {K.~S.}\ \bibnamefont
  {Novoselov}}, \bibinfo {author} {\bibfnamefont {V.~I.}\ \bibnamefont
  {Fal'ko}}, \bibinfo {author} {\bibfnamefont {L.}~\bibnamefont {Colombo}},
  \bibinfo {author} {\bibfnamefont {P.~R.}\ \bibnamefont {FGellert}}, \bibinfo
  {author} {\bibfnamefont {M.~G.}\ \bibnamefont {Schwab}}, \ and\ \bibinfo
  {author} {\bibfnamefont {K.}~\bibnamefont {Kim}},\ }\href@noop {} {\bibfield
  {journal} {\bibinfo  {journal} {Nature}\ }\textbf {\bibinfo {volume} {490}},\
  \bibinfo {pages} {192} (\bibinfo {year} {2012})}\BibitemShut {NoStop}%
\bibitem [{\citenamefont {Shikin}\ \emph {et~al.}(2000)\citenamefont {Shikin},
  \citenamefont {Prudnikova}, \citenamefont {Adamchuk}, \citenamefont
  {Moresco},\ and\ \citenamefont {Rieder}}]{Shikin2000}%
  \BibitemOpen
  \bibfield  {author} {\bibinfo {author} {\bibfnamefont {A.~M.}\ \bibnamefont
  {Shikin}}, \bibinfo {author} {\bibfnamefont {G.~V.}\ \bibnamefont
  {Prudnikova}}, \bibinfo {author} {\bibfnamefont {V.~K.}\ \bibnamefont
  {Adamchuk}}, \bibinfo {author} {\bibfnamefont {F.}~\bibnamefont {Moresco}}, \
  and\ \bibinfo {author} {\bibfnamefont {K.-H.}\ \bibnamefont {Rieder}},\
  }\href@noop {} {\bibfield  {journal} {\bibinfo  {journal} {Phys. Rev. B}\
  }\textbf {\bibinfo {volume} {62}},\ \bibinfo {pages} {13202} (\bibinfo {year}
  {2000})}\BibitemShut {NoStop}%
\bibitem [{\citenamefont {Dedkov}\ \emph {et~al.}(2001)\citenamefont {Dedkov},
  \citenamefont {Shikin}, \citenamefont {Adamchuk}, \citenamefont {Molodtsov},
  \citenamefont {Laubschat}, \citenamefont {Bauer},\ and\ \citenamefont
  {Kaindl}}]{Dedkov2001}%
  \BibitemOpen
  \bibfield  {author} {\bibinfo {author} {\bibfnamefont {Y.~S.}\ \bibnamefont
  {Dedkov}}, \bibinfo {author} {\bibfnamefont {A.~M.}\ \bibnamefont {Shikin}},
  \bibinfo {author} {\bibfnamefont {V.~K.}\ \bibnamefont {Adamchuk}}, \bibinfo
  {author} {\bibfnamefont {S.~L.}\ \bibnamefont {Molodtsov}}, \bibinfo {author}
  {\bibfnamefont {C.}~\bibnamefont {Laubschat}}, \bibinfo {author}
  {\bibfnamefont {A.}~\bibnamefont {Bauer}}, \ and\ \bibinfo {author}
  {\bibfnamefont {G.}~\bibnamefont {Kaindl}},\ }\href@noop {} {\bibfield
  {journal} {\bibinfo  {journal} {Phys. Rev. B}\ }\textbf {\bibinfo {volume}
  {64}},\ \bibinfo {pages} {035405} (\bibinfo {year} {2001})}\BibitemShut
  {NoStop}%
\bibitem [{\citenamefont {Riedl}\ \emph {et~al.}(2009)\citenamefont {Riedl},
  \citenamefont {Coletti}, \citenamefont {Iwasaki}, \citenamefont {Zakharov},\
  and\ \citenamefont {Starke}}]{Riedl2009}%
  \BibitemOpen
  \bibfield  {author} {\bibinfo {author} {\bibfnamefont {C.}~\bibnamefont
  {Riedl}}, \bibinfo {author} {\bibfnamefont {C.}~\bibnamefont {Coletti}},
  \bibinfo {author} {\bibfnamefont {T.}~\bibnamefont {Iwasaki}}, \bibinfo
  {author} {\bibfnamefont {A.~A.}\ \bibnamefont {Zakharov}}, \ and\ \bibinfo
  {author} {\bibfnamefont {U.}~\bibnamefont {Starke}},\ }\href@noop {}
  {\bibfield  {journal} {\bibinfo  {journal} {Phys. Rev. Lett.}\ }\textbf
  {\bibinfo {volume} {103}},\ \bibinfo {pages} {246804} (\bibinfo {year}
  {2009})}\BibitemShut {NoStop}%
\bibitem [{\citenamefont {Nagashima}\ \emph {et~al.}(1994)\citenamefont
  {Nagashima}, \citenamefont {Tejima},\ and\ \citenamefont
  {Oshima}}]{Nagashima1994}%
  \BibitemOpen
  \bibfield  {author} {\bibinfo {author} {\bibfnamefont {A.}~\bibnamefont
  {Nagashima}}, \bibinfo {author} {\bibfnamefont {N.}~\bibnamefont {Tejima}}, \
  and\ \bibinfo {author} {\bibfnamefont {C.}~\bibnamefont {Oshima}},\
  }\href@noop {} {\bibfield  {journal} {\bibinfo  {journal} {Phys. Rev. B}\
  }\textbf {\bibinfo {volume} {50}},\ \bibinfo {pages} {17487} (\bibinfo {year}
  {1994})}\BibitemShut {NoStop}%
\bibitem [{\citenamefont {Weser}\ \emph {et~al.}(2011)\citenamefont {Weser},
  \citenamefont {Voloshina}, \citenamefont {Horn},\ and\ \citenamefont
  {Dedkov}}]{Weser2011}%
  \BibitemOpen
  \bibfield  {author} {\bibinfo {author} {\bibfnamefont {M.}~\bibnamefont
  {Weser}}, \bibinfo {author} {\bibfnamefont {E.~N.}\ \bibnamefont
  {Voloshina}}, \bibinfo {author} {\bibfnamefont {K.}~\bibnamefont {Horn}}, \
  and\ \bibinfo {author} {\bibfnamefont {Y.~S.}\ \bibnamefont {Dedkov}},\
  }\href@noop {} {\bibfield  {journal} {\bibinfo  {journal} {Phys. Chem. Chem.
  Phys.}\ }\textbf {\bibinfo {volume} {13}},\ \bibinfo {pages} {7534} (\bibinfo
  {year} {2011})}\BibitemShut {NoStop}%
\bibitem [{\citenamefont {Rougemaille}\ \emph {et~al.}(2012)\citenamefont
  {Rougemaille}, \citenamefont {N'Diaye}, \citenamefont {Coraux}, \citenamefont
  {Vo-Van}, \citenamefont {Fruchart},\ and\ \citenamefont
  {Schmid}}]{Rougemaille2012}%
  \BibitemOpen
  \bibfield  {author} {\bibinfo {author} {\bibfnamefont {N.}~\bibnamefont
  {Rougemaille}}, \bibinfo {author} {\bibfnamefont {A.~T.}\ \bibnamefont
  {N'Diaye}}, \bibinfo {author} {\bibfnamefont {J.}~\bibnamefont {Coraux}},
  \bibinfo {author} {\bibfnamefont {C.}~\bibnamefont {Vo-Van}}, \bibinfo
  {author} {\bibfnamefont {O.}~\bibnamefont {Fruchart}}, \ and\ \bibinfo
  {author} {\bibfnamefont {A.~K.}\ \bibnamefont {Schmid}},\ }\href@noop {}
  {\bibfield  {journal} {\bibinfo  {journal} {Appl. Phys. Lett.}\ }\textbf
  {\bibinfo {volume} {101}},\ \bibinfo {pages} {142403} (\bibinfo {year}
  {2012})}\BibitemShut {NoStop}%
\bibitem [{\citenamefont {Decker}\ \emph {et~al.}(2013)\citenamefont {Decker},
  \citenamefont {Brede}, \citenamefont {Atodiresei}, \citenamefont {Caciuc},
  \citenamefont {Bl\"ugel},\ and\ \citenamefont {Wiesendanger}}]{Decker2013}%
  \BibitemOpen
  \bibfield  {author} {\bibinfo {author} {\bibfnamefont {R.}~\bibnamefont
  {Decker}}, \bibinfo {author} {\bibfnamefont {J.}~\bibnamefont {Brede}},
  \bibinfo {author} {\bibfnamefont {N.}~\bibnamefont {Atodiresei}}, \bibinfo
  {author} {\bibfnamefont {V.}~\bibnamefont {Caciuc}}, \bibinfo {author}
  {\bibfnamefont {S.}~\bibnamefont {Bl\"ugel}}, \ and\ \bibinfo {author}
  {\bibfnamefont {R.}~\bibnamefont {Wiesendanger}},\ }\href@noop {} {\bibfield
  {journal} {\bibinfo  {journal} {Phys. Rev. B}\ }\textbf {\bibinfo {volume}
  {87}},\ \bibinfo {pages} {041403} (\bibinfo {year} {2013})}\BibitemShut
  {NoStop}%
\bibitem [{\citenamefont {Sutter}\ \emph {et~al.}(2010)\citenamefont {Sutter},
  \citenamefont {Sadowski},\ and\ \citenamefont {Sutter}}]{Sutter2010}%
  \BibitemOpen
  \bibfield  {author} {\bibinfo {author} {\bibfnamefont {P.}~\bibnamefont
  {Sutter}}, \bibinfo {author} {\bibfnamefont {J.~T.}\ \bibnamefont
  {Sadowski}}, \ and\ \bibinfo {author} {\bibfnamefont {E.~A.}\ \bibnamefont
  {Sutter}},\ }\href@noop {} {\bibfield  {journal} {\bibinfo  {journal} {J. Am.
  Chem. Soc.}\ }\textbf {\bibinfo {volume} {132}},\ \bibinfo {pages} {8175}
  (\bibinfo {year} {2010})}\BibitemShut {NoStop}%
\bibitem [{\citenamefont {Starodub}\ \emph {et~al.}(2010)\citenamefont
  {Starodub}, \citenamefont {Bartelt},\ and\ \citenamefont
  {McCarty}}]{Starodub2010}%
  \BibitemOpen
  \bibfield  {author} {\bibinfo {author} {\bibfnamefont {E.}~\bibnamefont
  {Starodub}}, \bibinfo {author} {\bibfnamefont {N.~C.}\ \bibnamefont
  {Bartelt}}, \ and\ \bibinfo {author} {\bibfnamefont {K.~F.}\ \bibnamefont
  {McCarty}},\ }\href@noop {} {\bibfield  {journal} {\bibinfo  {journal} {J.
  Phys. Chem. C}\ }\textbf {\bibinfo {volume} {114}},\ \bibinfo {pages} {5134}
  (\bibinfo {year} {2010})}\BibitemShut {NoStop}%
\bibitem [{\citenamefont {Gr{\aa}n\"as}\ \emph {et~al.}(2012)\citenamefont
  {Gr{\aa}n\"as}, \citenamefont {Knudsen}, \citenamefont {Schröder},
  \citenamefont {Gerber}, \citenamefont {Busse}, \citenamefont {Arman},
  \citenamefont {Schulte}, \citenamefont {Andersen},\ and\ \citenamefont
  {Michely}}]{Granas2012}%
  \BibitemOpen
  \bibfield  {author} {\bibinfo {author} {\bibfnamefont {E.}~\bibnamefont
  {Gr{\aa}n\"as}}, \bibinfo {author} {\bibfnamefont {J.}~\bibnamefont
  {Knudsen}}, \bibinfo {author} {\bibfnamefont {U.~A.}\ \bibnamefont
  {Schröder}}, \bibinfo {author} {\bibfnamefont {T.}~\bibnamefont {Gerber}},
  \bibinfo {author} {\bibfnamefont {C.}~\bibnamefont {Busse}}, \bibinfo
  {author} {\bibfnamefont {M.~A.}\ \bibnamefont {Arman}}, \bibinfo {author}
  {\bibfnamefont {K.}~\bibnamefont {Schulte}}, \bibinfo {author} {\bibfnamefont
  {J.~N.}\ \bibnamefont {Andersen}}, \ and\ \bibinfo {author} {\bibfnamefont
  {T.}~\bibnamefont {Michely}},\ }\href@noop {} {\bibfield  {journal} {\bibinfo
   {journal} {ACS Nano}\ }\textbf {\bibinfo {volume} {6}},\ \bibinfo {pages}
  {9951} (\bibinfo {year} {2012})}\BibitemShut {NoStop}%
\bibitem [{\citenamefont {Wu}\ and\ \citenamefont {Ignatiev}(1983)}]{Wu1983}%
  \BibitemOpen
  \bibfield  {author} {\bibinfo {author} {\bibfnamefont {N.~J.}\ \bibnamefont
  {Wu}}\ and\ \bibinfo {author} {\bibfnamefont {A.}~\bibnamefont {Ignatiev}},\
  }\href@noop {} {\bibfield  {journal} {\bibinfo  {journal} {Phys. Rev. B}\
  }\textbf {\bibinfo {volume} {28}},\ \bibinfo {pages} {7288} (\bibinfo {year}
  {1983})}\BibitemShut {NoStop}%
\bibitem [{\citenamefont {Coraux}\ \emph {et~al.}(2012)\citenamefont {Coraux},
  \citenamefont {N’Diaye}, \citenamefont {Rougemaille}, \citenamefont
  {Vo-Van}, \citenamefont {Kimouche}, \citenamefont {Yang}, \citenamefont
  {Chshiev}, \citenamefont {Bendiab}, \citenamefont {Fruchart},\ and\
  \citenamefont {Schmid}}]{Coraux2012}%
  \BibitemOpen
  \bibfield  {author} {\bibinfo {author} {\bibfnamefont {J.}~\bibnamefont
  {Coraux}}, \bibinfo {author} {\bibfnamefont {A.~T.}\ \bibnamefont
  {N’Diaye}}, \bibinfo {author} {\bibfnamefont {N.}~\bibnamefont
  {Rougemaille}}, \bibinfo {author} {\bibfnamefont {C.}~\bibnamefont {Vo-Van}},
  \bibinfo {author} {\bibfnamefont {A.}~\bibnamefont {Kimouche}}, \bibinfo
  {author} {\bibfnamefont {H.-X.}\ \bibnamefont {Yang}}, \bibinfo {author}
  {\bibfnamefont {M.}~\bibnamefont {Chshiev}}, \bibinfo {author} {\bibfnamefont
  {N.}~\bibnamefont {Bendiab}}, \bibinfo {author} {\bibfnamefont
  {O.}~\bibnamefont {Fruchart}}, \ and\ \bibinfo {author} {\bibfnamefont
  {A.~K.}\ \bibnamefont {Schmid}},\ }\href@noop {} {\bibfield  {journal}
  {\bibinfo  {journal} {J. Phys. Chem. Lett.}\ }\textbf {\bibinfo {volume}
  {3}},\ \bibinfo {pages} {2059} (\bibinfo {year} {2012})}\BibitemShut
  {NoStop}%
\bibitem [{\citenamefont {Sicot}\ \emph {et~al.}(2012)\citenamefont {Sicot},
  \citenamefont {Leicht}, \citenamefont {Zusan}, \citenamefont {Bouvron},
  \citenamefont {Zander}, \citenamefont {Weser}, \citenamefont {Dedkov},
  \citenamefont {Horn},\ and\ \citenamefont {Fonin}}]{Sicot2012}%
  \BibitemOpen
  \bibfield  {author} {\bibinfo {author} {\bibfnamefont {M.}~\bibnamefont
  {Sicot}}, \bibinfo {author} {\bibfnamefont {P.}~\bibnamefont {Leicht}},
  \bibinfo {author} {\bibfnamefont {A.}~\bibnamefont {Zusan}}, \bibinfo
  {author} {\bibfnamefont {S.}~\bibnamefont {Bouvron}}, \bibinfo {author}
  {\bibfnamefont {O.}~\bibnamefont {Zander}}, \bibinfo {author} {\bibfnamefont
  {M.}~\bibnamefont {Weser}}, \bibinfo {author} {\bibfnamefont {Y.~S.}\
  \bibnamefont {Dedkov}}, \bibinfo {author} {\bibfnamefont {K.}~\bibnamefont
  {Horn}}, \ and\ \bibinfo {author} {\bibfnamefont {M.}~\bibnamefont {Fonin}},\
  }\href@noop {} {\bibfield  {journal} {\bibinfo  {journal} {ACS Nano}\
  }\textbf {\bibinfo {volume} {6}},\ \bibinfo {pages} {151} (\bibinfo {year}
  {2012})}\BibitemShut {NoStop}%
\bibitem [{\citenamefont {Huang}\ \emph {et~al.}(2011)\citenamefont {Huang},
  \citenamefont {Pan}, \citenamefont {Pan}, \citenamefont {Gao}, \citenamefont
  {Xu}, \citenamefont {Que}, \citenamefont {Zhou}, \citenamefont {Wang},
  \citenamefont {Du},\ and\ \citenamefont {Gao}}]{Huang2011}%
  \BibitemOpen
  \bibfield  {author} {\bibinfo {author} {\bibfnamefont {L.}~\bibnamefont
  {Huang}}, \bibinfo {author} {\bibfnamefont {Y.}~\bibnamefont {Pan}}, \bibinfo
  {author} {\bibfnamefont {L.}~\bibnamefont {Pan}}, \bibinfo {author}
  {\bibfnamefont {M.}~\bibnamefont {Gao}}, \bibinfo {author} {\bibfnamefont
  {W.}~\bibnamefont {Xu}}, \bibinfo {author} {\bibfnamefont {Y.}~\bibnamefont
  {Que}}, \bibinfo {author} {\bibfnamefont {H.}~\bibnamefont {Zhou}}, \bibinfo
  {author} {\bibfnamefont {Y.}~\bibnamefont {Wang}}, \bibinfo {author}
  {\bibfnamefont {S.}~\bibnamefont {Du}}, \ and\ \bibinfo {author}
  {\bibfnamefont {H.-J.}\ \bibnamefont {Gao}},\ }\href@noop {} {\bibfield
  {journal} {\bibinfo  {journal} {Appl. Phys. Lett.}\ }\textbf {\bibinfo
  {volume} {99}},\ \bibinfo {pages} {163107} (\bibinfo {year}
  {2011})}\BibitemShut {NoStop}%
\bibitem [{\citenamefont {Jin}\ \emph {et~al.}(2013)\citenamefont {Jin},
  \citenamefont {Fu}, \citenamefont {Yang},\ and\ \citenamefont
  {Bao}}]{Jin2013}%
  \BibitemOpen
  \bibfield  {author} {\bibinfo {author} {\bibfnamefont {L.}~\bibnamefont
  {Jin}}, \bibinfo {author} {\bibfnamefont {Q.}~\bibnamefont {Fu}}, \bibinfo
  {author} {\bibfnamefont {Y.}~\bibnamefont {Yang}}, \ and\ \bibinfo {author}
  {\bibfnamefont {X.}~\bibnamefont {Bao}},\ }\href@noop {} {\bibfield
  {journal} {\bibinfo  {journal} {Surf. Sci.}\ ,\ \bibinfo {pages} {in press}}
  (\bibinfo {year} {2013})}\BibitemShut {NoStop}%
\bibitem [{\citenamefont {Bauer}(1994)}]{Bauer1994}%
  \BibitemOpen
  \bibfield  {author} {\bibinfo {author} {\bibfnamefont {E.}~\bibnamefont
  {Bauer}},\ }\href@noop {} {\bibfield  {journal} {\bibinfo  {journal} {Rep.
  Progr. Phys.}\ }\textbf {\bibinfo {volume} {57}},\ \bibinfo {pages} {895}
  (\bibinfo {year} {1994})}\BibitemShut {NoStop}%
\bibitem [{\citenamefont {Altman}(2010)}]{Altman2010}%
  \BibitemOpen
  \bibfield  {author} {\bibinfo {author} {\bibfnamefont {M.~S.}\ \bibnamefont
  {Altman}},\ }\href@noop {} {\bibfield  {journal} {\bibinfo  {journal}
  {Journal of Physics: Condensed Matter}\ }\textbf {\bibinfo {volume} {22}},\
  \bibinfo {pages} {084017} (\bibinfo {year} {2010})}\BibitemShut {NoStop}%
\bibitem [{\citenamefont {N'Diaye}\ \emph {et~al.}(2009)\citenamefont
  {N'Diaye}, \citenamefont {van Gastel}, \citenamefont {Martínez-Galera},
  \citenamefont {Coraux}, \citenamefont {Hattab}, \citenamefont {Wall},
  \citenamefont {zu~Heringdorf}, \citenamefont {von Hoegen}, \citenamefont
  {Gómez-Rodríguez}, \citenamefont {Poelsema}, \citenamefont {Busse},\ and\
  \citenamefont {Michely}}]{Alpha2009}%
  \BibitemOpen
  \bibfield  {author} {\bibinfo {author} {\bibfnamefont {A.~T.}\ \bibnamefont
  {N'Diaye}}, \bibinfo {author} {\bibfnamefont {R.}~\bibnamefont {van Gastel}},
  \bibinfo {author} {\bibfnamefont {A.~J.}\ \bibnamefont {Martínez-Galera}},
  \bibinfo {author} {\bibfnamefont {J.}~\bibnamefont {Coraux}}, \bibinfo
  {author} {\bibfnamefont {H.}~\bibnamefont {Hattab}}, \bibinfo {author}
  {\bibfnamefont {D.}~\bibnamefont {Wall}}, \bibinfo {author} {\bibfnamefont
  {F.-J.~M.}\ \bibnamefont {zu~Heringdorf}}, \bibinfo {author} {\bibfnamefont
  {M.~H.}\ \bibnamefont {von Hoegen}}, \bibinfo {author} {\bibfnamefont
  {J.~M.}\ \bibnamefont {Gómez-Rodríguez}}, \bibinfo {author} {\bibfnamefont
  {B.}~\bibnamefont {Poelsema}}, \bibinfo {author} {\bibfnamefont
  {C.}~\bibnamefont {Busse}}, \ and\ \bibinfo {author} {\bibfnamefont
  {T.}~\bibnamefont {Michely}},\ }\href@noop {} {\bibfield  {journal} {\bibinfo
   {journal} {New J. Phys.}\ }\textbf {\bibinfo {volume} {11}},\ \bibinfo
  {pages} {113056} (\bibinfo {year} {2009})}\BibitemShut {NoStop}%
\bibitem [{\citenamefont {Hattab}\ \emph {et~al.}(2012)\citenamefont {Hattab},
  \citenamefont {N’Diaye}, \citenamefont {Wall}, \citenamefont {Klein},
  \citenamefont {Jnawali}, \citenamefont {Coraux}, \citenamefont {Busse},
  \citenamefont {van Gastel}, \citenamefont {Poelsema}, \citenamefont
  {Michely}, \citenamefont {Meyer~zu Heringdorf},\ and\ \citenamefont {Horn-von
  Hoegen}}]{Hattab2012}%
  \BibitemOpen
  \bibfield  {author} {\bibinfo {author} {\bibfnamefont {H.}~\bibnamefont
  {Hattab}}, \bibinfo {author} {\bibfnamefont {A.~T.}\ \bibnamefont
  {N’Diaye}}, \bibinfo {author} {\bibfnamefont {D.}~\bibnamefont {Wall}},
  \bibinfo {author} {\bibfnamefont {C.}~\bibnamefont {Klein}}, \bibinfo
  {author} {\bibfnamefont {G.}~\bibnamefont {Jnawali}}, \bibinfo {author}
  {\bibfnamefont {J.}~\bibnamefont {Coraux}}, \bibinfo {author} {\bibfnamefont
  {C.}~\bibnamefont {Busse}}, \bibinfo {author} {\bibfnamefont
  {R.}~\bibnamefont {van Gastel}}, \bibinfo {author} {\bibfnamefont
  {B.}~\bibnamefont {Poelsema}}, \bibinfo {author} {\bibfnamefont
  {T.}~\bibnamefont {Michely}}, \bibinfo {author} {\bibfnamefont {F.-J.}\
  \bibnamefont {Meyer~zu Heringdorf}}, \ and\ \bibinfo {author} {\bibfnamefont
  {M.}~\bibnamefont {Horn-von Hoegen}},\ }\href@noop {} {\bibfield  {journal}
  {\bibinfo  {journal} {Nano Letters}\ }\textbf {\bibinfo {volume} {12}},\
  \bibinfo {pages} {678} (\bibinfo {year} {2012})}\BibitemShut {NoStop}%
\bibitem [{\citenamefont {Locatelli}\ \emph {et~al.}(2006)\citenamefont
  {Locatelli}, \citenamefont {Aballe}, \citenamefont {Mente\c{s}},
  \citenamefont {Kiskinova},\ and\ \citenamefont {Bauer}}]{Locatelli2006}%
  \BibitemOpen
  \bibfield  {author} {\bibinfo {author} {\bibfnamefont {A.}~\bibnamefont
  {Locatelli}}, \bibinfo {author} {\bibfnamefont {L.}~\bibnamefont {Aballe}},
  \bibinfo {author} {\bibfnamefont {T.~O.}\ \bibnamefont {Mente\c{s}}},
  \bibinfo {author} {\bibfnamefont {M.}~\bibnamefont {Kiskinova}}, \ and\
  \bibinfo {author} {\bibfnamefont {E.}~\bibnamefont {Bauer}},\ }\href@noop {}
  {\bibfield  {journal} {\bibinfo  {journal} {Surf. Int. Analysis}\ }\textbf
  {\bibinfo {volume} {38}},\ \bibinfo {pages} {1554} (\bibinfo {year}
  {2006})}\BibitemShut {NoStop}%
\bibitem [{mon()}]{mono}%
  \BibitemOpen
  \href@noop {} {}\bibinfo {note} {A monolayer is defined as the first atomic
  layer of Co that grows pseudo- morphically on Ir(111).}\BibitemShut {Stop}%
\bibitem [{\citenamefont {Loginova}\ \emph
  {et~al.}(2009{\natexlab{a}})\citenamefont {Loginova}, \citenamefont {Nie},
  \citenamefont {Th\"urmer}, \citenamefont {Bartelt},\ and\ \citenamefont
  {McCarty}}]{Loginova2009}%
  \BibitemOpen
  \bibfield  {author} {\bibinfo {author} {\bibfnamefont {E.}~\bibnamefont
  {Loginova}}, \bibinfo {author} {\bibfnamefont {S.}~\bibnamefont {Nie}},
  \bibinfo {author} {\bibfnamefont {K.}~\bibnamefont {Th\"urmer}}, \bibinfo
  {author} {\bibfnamefont {N.~C.}\ \bibnamefont {Bartelt}}, \ and\ \bibinfo
  {author} {\bibfnamefont {K.~F.}\ \bibnamefont {McCarty}},\ }\href@noop {}
  {\bibfield  {journal} {\bibinfo  {journal} {Phys. Rev. B}\ }\textbf {\bibinfo
  {volume} {80}},\ \bibinfo {pages} {085430} (\bibinfo {year}
  {2009}{\natexlab{a}})}\BibitemShut {NoStop}%
\bibitem [{\citenamefont {Meng}\ \emph {et~al.}(2012)\citenamefont {Meng},
  \citenamefont {Wu}, \citenamefont {Zhang}, \citenamefont {Li}, \citenamefont
  {Du}, \citenamefont {Wang},\ and\ \citenamefont {Gao}}]{Meng2012}%
  \BibitemOpen
  \bibfield  {author} {\bibinfo {author} {\bibfnamefont {L.}~\bibnamefont
  {Meng}}, \bibinfo {author} {\bibfnamefont {R.}~\bibnamefont {Wu}}, \bibinfo
  {author} {\bibfnamefont {L.}~\bibnamefont {Zhang}}, \bibinfo {author}
  {\bibfnamefont {L.}~\bibnamefont {Li}}, \bibinfo {author} {\bibfnamefont
  {S.}~\bibnamefont {Du}}, \bibinfo {author} {\bibfnamefont {Y.}~\bibnamefont
  {Wang}}, \ and\ \bibinfo {author} {\bibfnamefont {H.-J.}\ \bibnamefont
  {Gao}},\ }\href@noop {} {\bibfield  {journal} {\bibinfo  {journal} {J. Phys.:
  Condens. Matter}\ }\textbf {\bibinfo {volume} {24}},\ \bibinfo {pages}
  {314214} (\bibinfo {year} {2012})}\BibitemShut {NoStop}%
\bibitem [{\citenamefont {Vo-Van}\ \emph {et~al.}(2011)\citenamefont {Vo-Van},
  \citenamefont {Schumacher}, \citenamefont {Coraux}, \citenamefont {Sessi},
  \citenamefont {Fruchart}, \citenamefont {Brookes}, \citenamefont {Ohresser},\
  and\ \citenamefont {Michely}}]{Vovan2011}%
  \BibitemOpen
  \bibfield  {author} {\bibinfo {author} {\bibfnamefont {C.}~\bibnamefont
  {Vo-Van}}, \bibinfo {author} {\bibfnamefont {S.}~\bibnamefont {Schumacher}},
  \bibinfo {author} {\bibfnamefont {J.}~\bibnamefont {Coraux}}, \bibinfo
  {author} {\bibfnamefont {V.}~\bibnamefont {Sessi}}, \bibinfo {author}
  {\bibfnamefont {O.}~\bibnamefont {Fruchart}}, \bibinfo {author}
  {\bibfnamefont {N.~B.}\ \bibnamefont {Brookes}}, \bibinfo {author}
  {\bibfnamefont {P.}~\bibnamefont {Ohresser}}, \ and\ \bibinfo {author}
  {\bibfnamefont {T.}~\bibnamefont {Michely}},\ }\href@noop {} {\bibfield
  {journal} {\bibinfo  {journal} {Appl. Phys. Lett.}\ }\textbf {\bibinfo
  {volume} {99}},\ \bibinfo {pages} {142504} (\bibinfo {year}
  {2011})}\BibitemShut {NoStop}%
\bibitem [{\citenamefont {Starodub}\ \emph {et~al.}(2011)\citenamefont
  {Starodub}, \citenamefont {Bostwick}, \citenamefont {Moreschini},
  \citenamefont {Nie}, \citenamefont {Gabaly}, \citenamefont {McCarty},\ and\
  \citenamefont {Rotenberg}}]{Starodub2011}%
  \BibitemOpen
  \bibfield  {author} {\bibinfo {author} {\bibfnamefont {E.}~\bibnamefont
  {Starodub}}, \bibinfo {author} {\bibfnamefont {A.}~\bibnamefont {Bostwick}},
  \bibinfo {author} {\bibfnamefont {L.}~\bibnamefont {Moreschini}}, \bibinfo
  {author} {\bibfnamefont {S.}~\bibnamefont {Nie}}, \bibinfo {author}
  {\bibfnamefont {F.~E.}\ \bibnamefont {Gabaly}}, \bibinfo {author}
  {\bibfnamefont {K.~F.}\ \bibnamefont {McCarty}}, \ and\ \bibinfo {author}
  {\bibfnamefont {E.}~\bibnamefont {Rotenberg}},\ }\href@noop {} {\bibfield
  {journal} {\bibinfo  {journal} {Phys. Rev. B}\ }\textbf {\bibinfo {volume}
  {83}},\ \bibinfo {pages} {125428} (\bibinfo {year} {2011})}\BibitemShut
  {NoStop}%
\bibitem [{\citenamefont {Giovannetti}\ \emph {et~al.}(2008)\citenamefont
  {Giovannetti}, \citenamefont {Khomyakov}, \citenamefont {Brocks},
  \citenamefont {Karpan}, \citenamefont {van~den Brink},\ and\ \citenamefont
  {Kelly}}]{Giovannetti2008}%
  \BibitemOpen
  \bibfield  {author} {\bibinfo {author} {\bibfnamefont {G.}~\bibnamefont
  {Giovannetti}}, \bibinfo {author} {\bibfnamefont {P.~A.}\ \bibnamefont
  {Khomyakov}}, \bibinfo {author} {\bibfnamefont {G.}~\bibnamefont {Brocks}},
  \bibinfo {author} {\bibfnamefont {V.~M.}\ \bibnamefont {Karpan}}, \bibinfo
  {author} {\bibfnamefont {J.}~\bibnamefont {van~den Brink}}, \ and\ \bibinfo
  {author} {\bibfnamefont {P.~J.}\ \bibnamefont {Kelly}},\ }\href@noop {}
  {\bibfield  {journal} {\bibinfo  {journal} {Phys. Rev. Lett.}\ }\textbf
  {\bibinfo {volume} {101}},\ \bibinfo {pages} {026803} (\bibinfo {year}
  {2008})}\BibitemShut {NoStop}%
\bibitem [{\citenamefont {Pacil\'e}\ \emph {et~al.}(2013)\citenamefont
  {Pacil\'e}, \citenamefont {Leicht}, \citenamefont {Papagno}, \citenamefont
  {Sheverdyaeva}, \citenamefont {Moras}, \citenamefont {Carbone}, \citenamefont
  {Krausert}, \citenamefont {Zielke}, \citenamefont {Fonin}, \citenamefont
  {Dedkov}, \citenamefont {Mittendorfer}, \citenamefont {Doppler},
  \citenamefont {Garhofer},\ and\ \citenamefont {Redinger}}]{Pacile2013}%
  \BibitemOpen
  \bibfield  {author} {\bibinfo {author} {\bibfnamefont {D.}~\bibnamefont
  {Pacil\'e}}, \bibinfo {author} {\bibfnamefont {P.}~\bibnamefont {Leicht}},
  \bibinfo {author} {\bibfnamefont {M.}~\bibnamefont {Papagno}}, \bibinfo
  {author} {\bibfnamefont {P.~M.}\ \bibnamefont {Sheverdyaeva}}, \bibinfo
  {author} {\bibfnamefont {P.}~\bibnamefont {Moras}}, \bibinfo {author}
  {\bibfnamefont {C.}~\bibnamefont {Carbone}}, \bibinfo {author} {\bibfnamefont
  {K.}~\bibnamefont {Krausert}}, \bibinfo {author} {\bibfnamefont
  {L.}~\bibnamefont {Zielke}}, \bibinfo {author} {\bibfnamefont
  {M.}~\bibnamefont {Fonin}}, \bibinfo {author} {\bibfnamefont {Y.~S.}\
  \bibnamefont {Dedkov}}, \bibinfo {author} {\bibfnamefont {F.}~\bibnamefont
  {Mittendorfer}}, \bibinfo {author} {\bibfnamefont {J.}~\bibnamefont
  {Doppler}}, \bibinfo {author} {\bibfnamefont {A.}~\bibnamefont {Garhofer}}, \
  and\ \bibinfo {author} {\bibfnamefont {J.}~\bibnamefont {Redinger}},\
  }\href@noop {} {\bibfield  {journal} {\bibinfo  {journal} {Phys. Rev. B}\
  }\textbf {\bibinfo {volume} {87}},\ \bibinfo {pages} {035420} (\bibinfo
  {year} {2013})}\BibitemShut {NoStop}%
\bibitem [{\citenamefont {Coraux}\ \emph {et~al.}(2008)\citenamefont {Coraux},
  \citenamefont {N'Diaye}, \citenamefont {Busse},\ and\ \citenamefont
  {Michely}}]{Coraux2008}%
  \BibitemOpen
  \bibfield  {author} {\bibinfo {author} {\bibfnamefont {J.}~\bibnamefont
  {Coraux}}, \bibinfo {author} {\bibfnamefont {A.~T.}\ \bibnamefont {N'Diaye}},
  \bibinfo {author} {\bibfnamefont {C.}~\bibnamefont {Busse}}, \ and\ \bibinfo
  {author} {\bibfnamefont {T.}~\bibnamefont {Michely}},\ }\href@noop {}
  {\bibfield  {journal} {\bibinfo  {journal} {Nano Letters}\ }\textbf {\bibinfo
  {volume} {8}},\ \bibinfo {pages} {565} (\bibinfo {year} {2008})}\BibitemShut
  {NoStop}%
\bibitem [{com()}]{comment}%
  \BibitemOpen
  \href@noop {} {}\bibinfo {note} {In fact, a corrugation is also expected at
  graphene grain boundaries. This issue has been addressed theoretically in Y.
  Liu and B. I. Yacobson, Nano Letters 10, 2178 (2010), where a radius of
  curvature of the order of 2-3 nm is predicted, i.e. much larger values than
  the ones expected at wrinkles or substrate step edges. To our knowledge,
  experimental determination of height variations accross pentagon-heptagon
  pairs is lacking.}\BibitemShut {Stop}%
\bibitem [{\citenamefont {Loginova}\ \emph
  {et~al.}(2009{\natexlab{b}})\citenamefont {Loginova}, \citenamefont
  {Bartelt}, \citenamefont {Feibelman},\ and\ \citenamefont
  {McCarty}}]{Loginova2009b}%
  \BibitemOpen
  \bibfield  {author} {\bibinfo {author} {\bibfnamefont {E.}~\bibnamefont
  {Loginova}}, \bibinfo {author} {\bibfnamefont {N.~C.}\ \bibnamefont
  {Bartelt}}, \bibinfo {author} {\bibfnamefont {P.~J.}\ \bibnamefont
  {Feibelman}}, \ and\ \bibinfo {author} {\bibfnamefont {K.~F.}\ \bibnamefont
  {McCarty}},\ }\href@noop {} {\bibfield  {journal} {\bibinfo  {journal} {New
  J. Phys.}\ }\textbf {\bibinfo {volume} {11}},\ \bibinfo {pages} {063046}
  (\bibinfo {year} {2009}{\natexlab{b}})}\BibitemShut {NoStop}%
\bibitem [{\citenamefont {see also in the case of graphene~on Ir(001)}\ \emph
  {et~al.}(2013)\citenamefont {see also in the case of graphene~on Ir(001)},
  \citenamefont {Locatelli}, \citenamefont {Wang}, \citenamefont {Africh},
  \citenamefont {Stoji\'{c}}, \citenamefont {Mente\c{s}}, \citenamefont
  {Comelli},\ and\ \citenamefont {Binggeli}}]{Locatelli2013}%
  \BibitemOpen
  \bibfield  {author} {\bibinfo {author} {\bibnamefont {see also in the case of
  graphene~on Ir(001)}}, \bibinfo {author} {\bibfnamefont {A.}~\bibnamefont
  {Locatelli}}, \bibinfo {author} {\bibfnamefont {C.}~\bibnamefont {Wang}},
  \bibinfo {author} {\bibfnamefont {C.}~\bibnamefont {Africh}}, \bibinfo
  {author} {\bibfnamefont {N.}~\bibnamefont {Stoji\'{c}}}, \bibinfo {author}
  {\bibfnamefont {T.~O.}\ \bibnamefont {Mente\c{s}}}, \bibinfo {author}
  {\bibfnamefont {G.}~\bibnamefont {Comelli}}, \ and\ \bibinfo {author}
  {\bibfnamefont {N.}~\bibnamefont {Binggeli}},\ }\href@noop {} {\bibfield
  {journal} {\bibinfo  {journal} {ACS Nano}\ }\textbf {\bibinfo {volume} {7}},\
  \bibinfo {pages} {6955} (\bibinfo {year} {2013})}\BibitemShut {NoStop}%
\bibitem [{\citenamefont {Thrower}\ and\ \citenamefont
  {Mayer}(1978)}]{Thrower1978}%
  \BibitemOpen
  \bibfield  {author} {\bibinfo {author} {\bibfnamefont {P.~A.}\ \bibnamefont
  {Thrower}}\ and\ \bibinfo {author} {\bibfnamefont {R.~M.}\ \bibnamefont
  {Mayer}},\ }\href@noop {} {\bibfield  {journal} {\bibinfo  {journal} {Physica
  Status Solidi (a)}\ }\textbf {\bibinfo {volume} {47}},\ \bibinfo {pages} {11}
  (\bibinfo {year} {1978})}\BibitemShut {NoStop}%
\bibitem [{\citenamefont {Boukhvalov}\ and\ \citenamefont
  {Katsnelson}(2009)}]{Boukhvalov2009}%
  \BibitemOpen
  \bibfield  {author} {\bibinfo {author} {\bibfnamefont {D.~W.}\ \bibnamefont
  {Boukhvalov}}\ and\ \bibinfo {author} {\bibfnamefont {M.~I.}\ \bibnamefont
  {Katsnelson}},\ }\href@noop {} {\bibfield  {journal} {\bibinfo  {journal}
  {Appl. Phys. Lett.}\ }\textbf {\bibinfo {volume} {95}},\ \bibinfo {pages}
  {023109} (\bibinfo {year} {2009})}\BibitemShut {NoStop}%
\bibitem [{\citenamefont {Krasheninnikov}\ \emph {et~al.}(2006)\citenamefont
  {Krasheninnikov}, \citenamefont {Lehtinen}, \citenamefont {Foster},\ and\
  \citenamefont {Nieminen}}]{Krasheninnikov2006}%
  \BibitemOpen
  \bibfield  {author} {\bibinfo {author} {\bibfnamefont {A.}~\bibnamefont
  {Krasheninnikov}}, \bibinfo {author} {\bibfnamefont {P.}~\bibnamefont
  {Lehtinen}}, \bibinfo {author} {\bibfnamefont {A.}~\bibnamefont {Foster}}, \
  and\ \bibinfo {author} {\bibfnamefont {R.}~\bibnamefont {Nieminen}},\
  }\href@noop {} {\bibfield  {journal} {\bibinfo  {journal} {Chem. Phys.
  Lett.}\ }\textbf {\bibinfo {volume} {418}},\ \bibinfo {pages} {132} (\bibinfo
  {year} {2006})}\BibitemShut {NoStop}%
\end{thebibliography}

%

\end{document}